\documentstyle[12pt]{article}
\newcommand{\be}{\begin{equation}}
\newcommand{\ee}{\end{equation}}
\newcommand{\ba}{\begin{array}}
\newcommand{\ea}{\end{array}}
\newcommand{\beqa}{\begin{eqnarray}}
\newcommand{\eeqa}{\end{eqnarray}}

\newcommand{\NP}[1]{Nucl. Phys.\ {\bf #1}\ }
\newcommand{\PL}[1]{Phys. Lett.\ {\bf #1}\ }

\newcommand{\PR}[1]{Phys. Rev.\ {\bf #1}\ }
\newcommand{\PRL}[1]{Phys. Rev. Lett.\ {\bf #1}\ }

\newcommand{\ZP}[1]{Z. Phys.\ {\bf #1}\ }

\newcommand{\A}{\alpha}
\newcommand{\B}{\beta}
\newcommand{\D}{\delta}
\newcommand{\DE}{\Delta}
\newcommand{\G}{\gamma}

\newcommand{\vep}{\varepsilon}

\newcommand{\lm}{\lambda}

\newcommand{\LRG}{$SU(2)_L\times SU(2)_R\times U(1)_{B-L}\,$}
\newcommand{\conj}[1]{\overline{#1}}
\newcommand{\WD}[1]{\Gamma_{#1}}
\newcommand{\ordo}[1]{{\cal O}(#1)}
\newcommand{\lsim}
{\;\raise0.3ex\hbox{$<$\kern-0.75em\raise-1.1ex\hbox{$\sim$}}\;}
\newcommand{\gsim}
{\;\raise0.3ex\hbox{$>$\kern-0.75em\raise-1.1ex\hbox{$\sim$}}\;}

\begin{document}

\begin{titlepage}
\mbox{}\hfill\makebox[4cm][l]{TURKU-FL-P12}\newline
\mbox{}\hfill\makebox[4cm][l]{hep-ph/9410227}
\vfill
\Large
\begin{center}
{\bf  Updated parameter limits of the left-right symmetric model}

\bigskip
\normalsize
{ Jukka  Sirkka\footnote{\normalsize sirkka@sara.cc.utu.fi}}\\[15pt]
{\it Department of  Physics, University of Turku, 20500 Turku, Finland}

{September 1994}
\bigskip
\vfill
\normalsize
{\it Submitted to Physics Letters B}
\vfill
{\bf Abstract}

\end{center}
\normalsize

Bounds of the neutral current sector parameters of the left-right
symmetric model are investigated taking into account the low-energy
data, LEP-data and CDF-result for the top mass $m_t=174\pm
10^{+13}_{-12}$. It is found that in the case of the minimal scalar
sector with a left- and a right-handed triplet and a bidoublet Higgses
the mass of the heavy neutral gauge boson $M_{Z'}$ should be  larger
than 1.2 TeV,  assuming equal left- and right-handed gauge couplings and
a negligible VEV of the left-handed triplet. For larger values of the
ratio $g_L/g_R$ smaller values of $M_{Z'}$ are allowed.

\end{titlepage}

\newpage

\normalsize
\setcounter{page}{2}
\setcounter{footnote}{0}

\noindent{\bf 1. Introduction. }
The left-right symmetric model (LR-model) with  the gauge group \LRG
\cite{LR} is a very appealing extension of the Standard Model. It has
several attractive features. In this model, parity is a symmetry of  the
lagrangian and it is broken only spontaneously due to the form of the
scalar potential providing a natural explanation for the parity
violation. Furthermore, the $U(1)$ generator has a physical
interpretation  as the $B-L$ quantum number. Finally, the
seesaw-mechanism can be realized and it leads to very small  Majorana
masses for the neutrinos which are mainly left-handed and large Majorana
masses for the neutrinos which are mainly right-handed. In addition to
the Standard Model particle content, there are heavy charged gauge boson
$W'$ and neutral gauge boson $Z'$ and three right-handed neutrinos which
form as mentioned, together with left-handed neutrinos the six Majorana
mass eigenstates.

The purpose of this paper is to update the parameter limits for the
LR-model using the latest LEP results, low-energy data and the
CDF-result for the top mass $m_t=174\pm 10^{+13}_{-12}$ GeV \cite{CDF}.
We shall do this in three cases. First, we do not specify the scalar
sector of the LR-model. In this case we have three fitting parameters:
the tree level correction $\DE\rho_0=M_W^2/(M_Z^2\cos^2\theta_w)-1$  to
the parameter $\rho=G_{NC}/G_{CC}$, which measures the relative strength
of the  neutral and charged current effective four fermion interactions
and is unity in the Standard Model at the tree level; the mixing angle
$\xi_0$  between $Z$ and $Z'$ and the mass $M_{Z'}$ of the $Z'$-boson.
As the second case we consider the minimal LR-model, with left- and
right-handed triplets $\DE_{L,R}$ and a bidoublet $\Phi$ in the scalar
sector. In ref. \cite{DGKO}  the most general scalar potential of the
minimal LR-model was studied\footnote{However, it was assumed that the
parameters of the scalar potential are real.}. It was shown that the
potential has a minimum with the see-saw relation
$v_Lv_R=\gamma(k_1^2+k_2^2)$, where $v_{L,R}$ and $k_i$ are VEV
parameters of the left- and right-handed triplets and bidoublet,
respectively, while $\gamma$ is a particular combination  of the scalar
potential parameters and $k_i$'s. By analysing the mass limits of
neutrinos it was further shown, abandoning the possibility of
fine-tuning the Yukawa couplings and the scalar potential parameters,
that, to avoid the need to fine-tune the parameter $\gamma$ very close
to zero,  the most natural possibility is to have $v_R\gsim 10^7 $GeV.
Another possibility, to have $v_R$ and thus $M_{Z'}$ in observable
range, is to look for a new symmetry to eliminate the relevant terms
from the scalar potential to guarantee that $\gamma=0$ without
fine-tuning. In both cases $v_L$ becomes negligible. Thus we assume that
the VEV of the left-handed triplet vanishes, $v_L=0$.  In this case the
parameter $\DE\rho_0$  can be expressed in terms of the mixing angle
$\zeta$ of the charged gauge bosons and the ratio $M_Z^2/M_{Z'}^2$ and
the angle $\xi_0$ can be expressed in terms of the ratio
$M_Z^2/M_{Z'}^2$, leaving us with two fitting parameters. Finally we
assume that the angle $\zeta$ is negligibly small and perform the
data-analysis only with $M_{Z'}$ as the fitting parameter.

The present study differs from the previous ones in the
respect that we use the experimental value of the top mass as  a
constraint and that we study also the case where the gauge couplings
$g_L$ and $g_R$ corresponding to the subgroups $SU(2)_L$ and $SU(2)_R$
may differ  by performing the analysis with various values of the ratio
$\lm\equiv g_L/g_R$.  The motivation for doing this is that if the
LR-model is embedded in a grand unified theory, it can happen that the
discrete left-right symmetry is broken at much higher energy scale than
the weak scale, allowing $g_R\neq g_L$ in the low-energy phenomena.
For example, in the case of supersymmetric version of $SO(10)$ grand
unified theory, a chain of symmetry breakings can be realized that leads
to a \LRG breaking scale $\approx 1$ TeV and to a value of $\lm$ as
large as $1.2$\cite{DKR}.

\noindent{\bf 2. Basic structure of the LR-model.}
In the LR-model, with the gauge group \LRG, the left-handed leptons
$\psi_L=(\nu,l)_L$ are in the representation $({\bf 2,1},-1)$ and
right-handed ones $\psi_R =(\nu,l)_R$ in the representation $({\bf
1,2},-1)$. The quark sector is assigned correspondingly. In the minimal
LR-model the scalar sector contains fields $\Phi$, $\Delta_L$ and
$\Delta_R$   assigned to the representations $({\bf 2,2},0)$, $({\bf
3,1},2)$ and $({\bf 1,3},2)$, respectively. The vacuum expectation
values of the fields are
\be
\langle\Phi\rangle=\left(\begin{array}{cc}
k_1 & 0 \\ 0 & k_2 \end{array}\right),
\hspace{0.1cm}
\langle\Delta_L\rangle=\left(\begin{array}{c}
0 \\ 0 \\ v_L \end{array}\right),\hspace{0.1cm}
\langle\Delta_R\rangle=\left(\begin{array}{c}
0 \\ 0 \\ v_R
\end{array}\right).
\ee
As discussed in the Introduction, we shall set $v_L=0$. Due to these
VEV's, the group \LRG is broken down to the electromagnetic group
$U(1)_Q$ and six gauge bosons $W^{\pm}, W^{'\pm}, Z, Z'$ acquire mass.
The masses of the charged gauge bosons $W$ and $W'$ are found to be, in
the limit $v_R^2\gg k_1^2+k_2^2$,
\beqa
M_{W}^2 &=& g_L^2\bar{k}^2\left(1-\frac{k_1^2k_2^2}{\bar k^2v_R^2}\right),
\nonumber \\
M_{W'}^2 &=& g_R^2v_R^2,\label{mw}
\eeqa
where $\bar k^2=(k_1^2+k_2^2)/2$.   The masses of the neutral gauge
bosons  $Z, Z'$ read as
\beqa
M_{Z}^2 &=& \frac{g_L^2\bar k^2}{c_w^2}\left(1
-\frac{y^4\bar k^2}{2c_w^4v_R^2}\right),
\nonumber \\
M_{Z'}^2 &=& \frac{2c_w^2g_R^2v_R^2}{y^2},\label{mz}
\eeqa
where shorthand notation $c_w=\cos\theta_W$ for weak mixing angle has
been used. In the LR-model the weak mixing angle is defined through
\be
g_Ls_w=g'y=e.
\ee
Here $g'$ is the $U(1)_{B-L}$ gauge coupling and
\be
y=\sqrt{c_w^2-\lm^2s_w^2},
\ee
and $\lm=g_L/g_R$.

Using Eqs.
(\ref{mw}) and (\ref{mz}) one deduces the value of
the parameter $\Delta\rho_0$,
\be
\Delta\rho_0 = \frac{y^2\B}{\lm^2}-\frac{y^2\zeta^2}{2c_w^4\B}
\label{lrdrho},
\ee
where $\B=M_Z^2/M_{Z'}^2$. The $W$-$W'$ mixing angle $\zeta$ is  in the
minimal LR-model given by
\be
\zeta=\lm\frac{k_1k_2}{v_R^2}.
\ee
One should notice that $\Delta\rho_0$ can be either positive or
negative depending on the values of $\B$ and $\zeta$.

The neutral current lagrangian reads
\be
{\cal L}_{NC}=g_Lj_{3L}\cdot W_{3L}+g_Rj_{3R}\cdot W_{3R}
+g'J_{B-L}\cdot B,\label{nco}
\ee
where $W_{3L,3R}$ are the neutral $SU(2)_{L,R}$ gauge bosons, and $B$
is the gauge boson of $U(1)_{B-L}$. The fermion neutral currents have
a form
\be
j^\mu_{L,R}=\conj{\psi}\G^\mu T_{3L,3R}\psi,\hspace{0.5cm}
j^\mu_{B-L}=\conj{\psi}\G^\mu \frac 12(B-L)\psi.
\ee
The lagrangian (\ref{nco}) can be expressed in terms of the photon field
$A$ and the fields $Z_L$ and $Z_R$ requiring that photon
couples only to the electromagnetic current
$j_{em}=j_{3L}+j_{3R}+j_{B-L}$ and defining $Z_R$ to be that combination
of $W_{3L}$, $W_{3R}$ and $B$ that does not couple to $j_{3L}$.
It follows that $Z_L$ and $Z_R$ couple to the currents
$e/(s_wc_w)(j_{3L}-s_w^2j_{em})$ and $e/(s_wc_w\lm y)(y^2j_{3R}-
\lm^2s_w^2j_{B-L})$, respectively. After a rotation to the mass
eigenstate basis $Z,Z'$,
\be
\left(\begin{array}{c} Z \\ Z' \end{array}\right)=
\left(\begin{array}{cc}
\cos\xi_0 & \sin\xi_0 \\ -\sin\xi_0 & \cos\xi_0 \end{array}\right)
\left(\begin{array}{c} Z_L \\ Z_R \end{array}\right),
\ee
we can express the neutral current lagrangian in terms of the mass
eigenstates $A$, $Z$ and $Z'$. The mixing angle $\xi_0$ measures the
deviations of the $Z$-boson LR-model couplings  from the Standard Model
couplings. Since the Standard Model is tested to be valid with a good
accuracy, we can expand the $Z$-coupling in linear order and, at energy
scales much lower than $M_{Z'}$,  $Z'$-coupling in zeroth order in
$\xi_0$. The neutral current lagrangian then reads
\beqa
{\cal L}_{NC}=eA\cdot j_{em}&+&\frac{e}{s_wc_w}Z\cdot
\left[j_{3L}\left(1+\frac{s_w^2\lm}y\xi_0\right)-s_w^2j_{em}
\left(1+\frac{\lm}y\xi_0\right)
+j_{3R}\frac{c_w^2}{\lm y}\xi_0\right] \nonumber \\
&+&\frac{e}{s_wc_w}Z'\cdot
\left[j_{3L}\frac{s_w^2\lm}y-s_w^2j_{em}\frac{\lm}y
+j_{3R}\frac{c_w^2}{\lm y}\right].\label{ncf}
\eeqa
In the minimal LR-model $\xi_0$ reads, in the limit $M_{Z'}\gg M_Z$,
\be
\xi_0=\frac y{\lm}\B.
\ee

\noindent{\bf 3. The LR-model formulas for the observables.}
In Standard Model, the analyses of the  low-energy data are based on the
effective lagrangian of the form
\beqa
{\cal L}_{eff} &=& \frac{e^2(q^2)}{q^2}j_{em}(1)\cdot j_{em}(2)\nonumber \\
& &+4\sqrt{2}G_F\rho(q^2)[j_{3L}(1)-s^2_{eff}(q^2)j_{em}(1)]\cdot
[j_{3L}(2)-s^2_{eff}(q^2)j_{em}(2)].
\eeqa
Here the loop corrections are collected to form three effective
quantities $e^2(q^2)$, $\rho(q^2)$ and $s^2_{eff}(q^2)$, which depend on
the energy scale $\sqrt{|q^2|}$, such that ${\cal L}_{eff}$ preserves the form
of the tree level lagrangian. This can be naturally done also in the
context of the LR-model. However, one might wonder if the form of the
effective quantities $e^2$, $\rho$ and $s^2_{eff}$ is changed when
the tree level LR-model corrections are taken into account. It was shown in
\cite{MS} that, in leading order in quantities $\B$, $\xi_0$ and
$\DE\rho_0$ the changes can be parametrized with $\DE\rho_0$ only:
\beqa
\rho &=& 1+\DE\rho_{SM}+\DE\rho_0 \nonumber \\
s^2_{eff} &=& s^2(1+\DE\kappa_{SM})+c^2\DE\rho_0
\nonumber \\
e^2 &=& e^2_{SM},\label{lrpars}
\eeqa
where $\DE\rho_{SM}$ and $\DE\kappa_{SM}$ represent the Standard Model
loop corrections and
$c^2=1-s^2=M_W^2/M_Z^2$. Thus the low-energy lagrangian for
the LR-model can be written in the form
\beqa
{\cal L}_{eff} &=& \frac{e^2}{q^2}j_{em}(1)\cdot j_{em}(2)
\nonumber \\
& &+4\sqrt{2}G_F\rho \left[j_{3L}(1)\left(1+\frac{s^2\lm}y\xi_0\right)
-s^2_{eff}j_{em}(1)\left(1+\frac{\lm}y\xi_0\right)+j_{3R}(1)
\frac{c^2}{\lm y}\xi_0\right]\times
\nonumber  \\
& & \phantom{+4\sqrt{2}G_F\rho}\left[j_{3L}(2)\left(1+\frac{s^2\lm}y\xi_0
\right)
-s^2_{eff}j_{em}(2)\left(1+\frac{\lm}y\xi_0\right)+j_{3R}(2)\frac{c^2}{\lm y}
\xi_0\right]
\nonumber \\
& &+4\sqrt{2}G_F\B \left[j_{3L}(1)\frac{s^2\lm}y-s^2j_{em}(1)\frac{\lm}y
+j_{3R}(1)\frac{c^2}{\lm y}\right]\times
\nonumber \\
& &\phantom{+4\sqrt{2}G_F\B }\left[j_{3L}(2)\frac{s^2\lm}y
-s^2j_{em}(2)\frac{\lm}y+j_{3R}(2)\frac{c^2}{\lm y}\right].\label{lrnc}
\eeqa
Strictly speaking, the Eq. (\ref{lrpars}) for $s^2_{eff}$ is only valid
when $M_W$ and hence $s^2$ is used as an input. The parameter $s^2$ is
calculable  as a function of the other more presicely measured
parameters from the expression for the Fermi coupling constant,
which reads, when taking into account
the LR-model corrections,
\be
\frac1{\sqrt{2}}G_F=\frac{\pi\alpha}{2s^2c^2M_Z^2}(
1+\DE r -\frac{c^2}{s^2}\DE\rho_0+\D_F).\label{sirlin}
\ee
Here $\DE r$ represents the Standard Model loop corrections and $\D_F$
the LR-model tree level corrections to the muon decay rate. As $\D_F$ is
a second order correction in the parameters $\zeta$ and
$M_W^2/M_{W'}^2$, it will be  neglected in the following. By calculating
$s^2$ from Eq. (\ref{sirlin}) and substituting the result to Eq.
(\ref{lrpars}), one then obtains
\be
s^2_{eff}=s^2_{eff,SM}-\frac{s^2c^2}{c^2-s^2}\DE\rho_0.
\ee
In the relation (\ref{sirlin}), we have included in addition to the
$\ordo{\A}$-corrections also the
$\ordo{\A\A_s}$-corrections \cite{KH}
whereas in Eqs. (\ref{lrpars}) only $\ordo{\A}$-corrections \cite{MaSi} are
included. This is because the parameter $s^2$, calculated from relation
(\ref{sirlin}), enters also in the expressions of the LEP-observables,
which are measured with a much greater accuracy than the low-energy
observables.

{}From (\ref{lrnc}) one can write the model independent low-energy
parameters, as defined through the model independent effective
lagrangians, in terms of LR-model parameters. For deep inelastic
neutrino-hadron scattering the parameters $\vep_{L,R}(q)$ are defined
through the lagrangian
\be {\cal L}^{\nu
H}=\sqrt{2}G_F\conj{\nu}_L\G^\mu\nu_L\sum_q( \vep_L(q)\conj{q}_L\G_\mu
q_L+\vep_R(q)\conj{q}_R\G_\mu q_R), \label{nudis}
\ee
with the LR-model expressions
\beqa
\vep_L(q) &=&
\rho(1+A_L\xi_0)\left[T_3(1+A_L\xi_0+A_L^2\B)-Qs^2_{eff}
(1+A_Q\xi_0+A_LA_Q\B)\right], \nonumber \\ \vep_R(q) &=&
\rho(1+A_L\xi_0)\left[T_3(A_R\xi_0+A_LA_R\B)-Qs^2_{eff}
(1+A_Q\xi_0+A_LA_Q\B)\right],
\eeqa
where $A_L=s^2\lm /y$, $A_Q=\lm /y$ and $A_R=c^2/ \lm y$ and $T_3\equiv
T_{3L}=T_{3R}$. Note that in Eq. (\ref{nudis}), neutrinos are assumed to
be left-handed. But in the LR-model, neutrinos are most naturally
Majorana particles. The see-saw mechanism produces three heavy and three
light mass eigenstates, with mass matrices $M_N\approx v_Rh_M$ and
$M_\nu\approx M_DM_N^{-1}M_D^T$, respectively \cite{MohPal}.  Here $h_M$
is the matrix of Yukawa  couplings between leptons and right-handed
triplet scalar $\DE_R$ and $M_D=Fk_1+Gk_2$ is a Dirac mass term with
Yukawa coupling matrices $F$ and $G$. Further, the charged lepton mass
matrix has a form $M_l=Fk_2+Gk_1$. Assuming that neither of the two
terms in $M_l$ is negligible and neglecting the inter-generational
mixings between neutrinos, we have the see-saw relation between the
light and heavy neutrino masses
\be
m_\nu\approx \frac{m_l^2}{m_N}.
\ee
This implies, together with the experimental limits of the light neutrino
masses \cite{PDB},  $m_{\nu_1}<7.3$ eV, $m_{\nu_2}<0.27$ MeV and
$m_{\nu_3}<35$ MeV, approximate lower bounds for the heavy neutrinos:
\be
m_{N_1}\gsim 4 \mbox{\ GeV},\hspace{0.5cm}
m_{N_2}\gsim 40 \mbox{\ GeV},\hspace{0.5cm}
m_{N_3}\gsim 90 \mbox{\ GeV}.\label{Nmasses}
\ee
Further, the current eigenstates $\nu_L$ and $\nu_R$ can be expressed in
terms of the mass eigenstates $\chi$ through a unitary transformation,
\beqa
\nu_{Li}=U_{Ll,ij}\chi_{l,j}+U_{Lh,ij}\chi_{h,j},\nonumber \\
\nu_{Ri}=U^*_{Rl,ij}\chi_{l,j}+U^*_{Rh,ij}\chi_{h,j},\label{mix}
\eeqa
where $U_{Ll}$ etc. are $3\times 3$ submatrices of a unitary $6\times 6$
matrix $U$ and $\chi_h$,
$\chi_l$ denote the heavy and light Majorana neutrinos, respectively.
The see-saw mechanism  implies that $U_{Lh}$ and $U_{Rl}$
are $\ordo{m_l/m_N}$ and $U_{Ll}$ and $U_{Rh}$ are $\ordo{1}$ \cite{MohPal}.
We can now write the left- and right-handed parts of the neutrino neutral
current effectively as
\beqa
\conj{\nu}_L\G^\mu\nu_L &=& \conj{\chi}_{lL}\G^\mu U_{Ll}^{\dagger}U_{Ll}
\chi_{lL}
+\dots=\conj{\chi}_{lL}\G^\mu\chi_{lL}+O(m_l^2/m_N^2)+\dots
\nonumber \\
\conj{\nu}_R\G^\mu\nu_R &=& \conj{\chi}_{lR}\G^\mu
U_{Rl}^TU_{Rl}^*\chi_{lR}
+\dots=(m_l^2/m_N^2)+\dots,
\eeqa
where dots represent the contribution where there is at least one heavy
neutrino involved. When the limits (\ref{Nmasses}) apply, the production
of heavy neutrinos is forbidden at low-energy scales and the lagrangian
(\ref{nudis}) is applicable.

Note also that, as the parameters $\vep_L$ and $\vep_R$ are  determined
from the ratios $R=\sigma_{\nu N}^{NC}/\sigma_{\nu N}^{CC}$ and
$\bar{R}=\sigma_{\bar{\nu} N}^{NC}/\sigma_{\bar{\nu} N}^{CC}$ of the
neutral and charged current cross sections of deep inelastic neutrino
scattering, one needs in principle to consider also the  charged sector
of the LR-model. However, it is straightforward to check that this
contribution is of the second order in parameters $\zeta$ and
$M_W^2/M_{W'}^2$ and as such, negligible.

The effective lagrangian for the electron-neutrino scattering defines
the parameters $g_V^e$ and $g_A^e$ according to
\be
{\cal L}^{\nu e}=\sqrt{2}G_F\conj{\nu}_L\G^\mu\nu_L
\conj{e}\G_\mu(g^e_V-g^e_A)e.
\ee
The LR-model expressions for them are
\beqa
g^e_V &=& \rho(1+A_L\xi_0)\left(-\frac 12(1+(A_L+A_R)\xi_0+A_L(A_L+A_R)\B)
\right.
\nonumber \\
& & \left. +2s^2_{eff}(1+A_Q\xi_0+A_LA_Q\B)\right), \nonumber \\
g^e_A &=& -\frac 12\rho\left(1+A_L\xi_0)(1+(A_L-A_R)\xi_0+A_L(A_L-A_R)\B
\right).
\eeqa
For the $\nu_e$-$e$ scattering the charged current contribution must be
included. Again, it is easy to check that the charged current LR-model
contribution to the cross-section is a negligible second order term in
parameters $\xi_0$ and $M_W^2/M_{W'}^2$.

The effective parity violating lagrangian in the electron-hadron
scattering defines the parameters $C_{iq}$ according to
\be
{\cal L}^{eH}=\frac{G_F}{\sqrt{2}}\sum_q(C_{1q}\conj{e}\G_\mu\G^5e
\conj{q}\G^\mu q+C_{2q}\conj{e}\G_\mu e\conj{q}\G^\mu\G^5q),
\ee
with the LR-model expressions
\beqa
C_{1q} &=& \rho\left(1+(A_L-A_R)\xi_0\right)\left(-T_{3q}(1+(A_L+A_R)\xi_0
+(A_L^2-A_R^2)\B)\right.
\nonumber \\
& & \left. +2s^2_{eff}Q_q(1+A_Q\xi_0+A_Q(A_L-A_R)\B)\right), \nonumber \\
C_{2q} &=& 2T_{3q}\rho\left(1+(A_L-A_R)\xi_0\right)\left(-\frac 12
(1+(A_L+A_R)\xi_0
+(A_L^2-A_R^2)\B)\right.
\nonumber \\
& & \left. +2s^2_{eff}(1+A_Q\xi_0+A_Q(A_L-A_R)\B)\right).
\eeqa
The parameters $\rho$ and $s^2_{eff}$ in the low-energy formulas depend
slightly on the process in question. Furthermore, there are some
additional terms from the box graphs \cite{MaSi,PDBL}, which should be
included.
The experimental values of the lowenergy parameters are taken
from Ref. \cite{PDBL}.

In the $Z$-line shape measurement at LEP, the $e\conj{e}\rightarrow
f\conj{f}(\gamma)$ cross-sections are fitted, after subtracting the pure
QED effects and the $\gamma$-$Z$ interference term, to the function
\be
\sigma^0(s)=\sigma^p_f \frac{s\WD{Z}^2}{(s-M_Z^2)^2+s^2\WD{Z}^2/M_Z^2},
\ee
where
\be
\sigma^p_f=\frac{12\pi\WD{e}\WD{f}}{M_Z^2\WD{Z}^2}.
\ee
An additional gauge boson would give a contribution to the cross-section
\cite{BCMS}
\be
\frac{\D\sigma_0}{\sigma_0}\approx\D R_{ZZ'}\frac{s-M_Z^2}{M_Z^2},
\label{dsigma}
\ee
where
\be
\D R_{ZZ'}\approx -2\frac{M_Z^2}{M_{Z'}^2}\frac{v_ev'_e+a_ea'_e}{v_e^2+a_e^2}
\frac{v_fv'_f+a_fa'_f}{v_f^2+a_f^2}\label{R},
\ee
where $v_f$ and $a_f$ are vector and axial-vector couplings of the
fermion $f$ to the $Z$-boson while $v'_f$ and $a'_f$ are the corresponding
quantities for $Z'$. Presence of the term (\ref{dsigma})
could in principle affect the
line shape parameters, but it turns out that this effect is negligible
even for modest values of $M_{Z'}$. For example, the
location of the maximum of the cross-section gets shifted by an amount
\be
\frac{\D s_0}{s_0}\approx \frac{\WD{Z}^2}{2M_Z^2}\D
R_{ZZ'}\approx 4\cdot 10^{-4} \D R_{ZZ'}.
\ee
Using $\D s_0/s_0\approx 2\D M_Z/M_Z$ and $M_Z=(91.1899\pm
0.0044)$ GeV \cite{GLASGOW} and taking the couplings in (\ref{R}) to be
equal for $Z$ and $Z'$ requires $M_Z'\lsim 300$ GeV for the additional
gauge boson to give a measurable contribution.

Thus the LEP measurements are sensitive only to the  parameters
$\DE\rho_0$ and $\xi_0$ through the dependence of the couplings  $v_f$
and $a_f$ of them. The form of the couplings can be read from the
lagrangian (\ref{ncf}) by replacing the bare quantity $s_w^2$ with the
effective quantity $s^2_f$.

We shall use the following high energy observables in the analysis:
the total width of the $Z$-boson $\WD{Z}$, the hadronic peak
cross-section $\sigma_p^{had}$, the ratio $R_l$ between the hadronic and
leptonic widths and the mass of the $Z$, the ratio $R_b$ between the
partial width to a $b\conj{b}$-pair and the hadronic width,
the mass of the $Z$
and the effective leptonic weak mixing angle defined through
\be
\sin^2\theta_w^{eff}=\frac 14(1-\frac{v_l}{a_l}),
\ee
which can be extracted from any of the leptonic asymmetries $A_{FB}$,
$P_\tau$, $A_{FB}^{pol(\tau)}$ or $A_{LR}$.
%For the line shape measurement of $Z$,
%assuming universality of the lepton couplings, there are four
%measurable quantities, which can be conveniently chosen to
%be the total width of the $Z$-boson $\WD{Z}$, the hadronic peak
%cross-section $\sigma_p^{had}$, the ratio $R_l$ between the hadronic and
%leptonic widths and the mass of the $Z$. In addition to them, we use in
%the analysis the effective leptonic  weak mixing angle defined by
%\be
%\sin^2\theta_w^{eff}=\frac 14(1-\frac{v_l}{a_l}),
%\ee
%which can be extracted from any of the leptonic asymmetries $A_{FB}$,
%$P_\tau$, $A_{FB}^{pol(\tau)}$ or $A_{LR}$.
Using mass of the $Z$ as
input leaves us five observables, of which $\WD{Z}$, $\sigma_p^{had}$,
$R_l$ and $R_b$ can be expressed in terms of the partial fermionic widths.
The widths have the form
\be
\WD{f}=\frac{G_F M_Z^3 \rho_f}{6\sqrt{2}\pi}(v_f^2+a_f^2)
(1+\frac{3\A}{4\pi}Q_f^2)K_{QCD}\label{width},
\ee
where
\beqa
v_f &=& T_{3f}(1+(A_L+A_R)\xi_0)-2s_f^2Q_f(1+A_Q\xi_0), \nonumber \\
a_f &=& T_{3f}(1+(A_L-A_R)\xi_0) \label{vecax}
\eeqa
and the QCD correction factor is defined by
\beqa
K_{QCD} &=& 3(1+\frac{\A_s}{\pi})\mbox{\hspace{0.5cm} for quarks}\nonumber \\
        &=& 1\phantom{(1+\frac{\A_s}{\pi})}\mbox{\hspace{0.5cm} for leptons}.
\eeqa
The partial width to a $b\bar{b}$-pair has a slightly different
behaviour due to the large contribution from the $Zb\bar{b}$-vertex.
This is taken into account by a parameter $\D_{vb}$ defined through
\be
\WD{b}=\WD{d}(1+\D_{vb})\label{bdif}.
\ee
In the limit of the large top mass it has the form \cite{BV}
\be
\D_{vb}=-\frac{20}{13}\frac{\A}{\pi}(\frac{m_t^2}{M_Z^2}+\frac{13}6
\ln\frac{m_t^2}{M_Z^2}).
\ee

The Eqs. (\ref{width}) and (\ref{vecax}) can also be applied to the case
of light neutrinos after removing the $A_R\xi_0$ terms. The partial
widths to a light and a heavy neutrino and to a heavy neutrino pair can
be neglected even if these decays are kinematically allowed.
This is because the widths are proportional to
\be
\left|g_L(\chi_i\chi_j)\right|^2+\left|g_R(\chi_i\chi_j)\right|^2,
\ee
where $g_{L,R}(\chi_i\chi_j)$ are left- and right- handed couplings of the
$Z$ to the neutrinos $\chi_i$ and $\chi_j$. By substituting Eq.
(\ref{mix}) to the lagrangian (\ref{ncf}) one deduces that the couplings
$g_{L,R}(\chi_i\chi_j)$, except the left-handed couplings of the light
neutrinos, are proportional at least to the first power of the
parameters $\xi_0$ or $U_{Lh}=\ordo{m_l/m_N}$ and hence give a negligible
second order contribution to the partial widths.

The quantities $\rho_f$ and $s^2_f$
have the same dependence on the parameter $\DE\rho_0$ as the
corresponding low-energy quantities. The Standard Model loop corrections
for them differ slightly because of the different energy scale and the
non-negligible vertex corrections. In
addition to the $\ordo{\A}$-corrections \cite{MaSi}, we have also included
$\ordo{\A\A_s}$-corrections  \cite{KH} in the expressions of
the parameters $\rho_f$ and $s^2_f$.   Note that $\sin^2\theta_w^{eff}$
is equal to $s_l^2$ in the absence of LR-corrections.
The values of the high energy observables to be used in our analysis
are \cite{GLASGOW}
\beqa
\Gamma_Z &=& 2.4974\pm 0.0038 \mbox{\ GeV}, \nonumber \\
\sigma_p^{had} &=& 41.49\pm 0.12 \mbox{\ nb}, \nonumber \\
R_l &=& 20.795 \pm 0.040, \nonumber \\
R_b &=& 0.2192 \pm 0.0018, \nonumber \\
\sin^2\theta_w^{eff} &=& 0.2317 \pm 0.0004.
\eeqa
For the quantities $\Gamma_Z$, $\sigma_p^{had}$ and $R_l$ we have
applied the
correlations used by the DELPHI Collaboration \cite{DELPHI} (i.e.
$c_{12}=-0.20$,
$c_{13}=0.00$ and $c_{23}=0.14$).

In addition to the low-energy and LEP-data we also use the $W$-mass
value $M_W=80.23\pm 0.18$ \cite{GLASGOW} as constraint, theoretical value
for $M_W$ being calculable from Eq. (\ref{sirlin}).

\noindent{\bf 4. Results and discussion.}  We have performed a
$\chi^2$-function minimization to fit the LR-model parameters with
various values of $\lm=g_L/g_R$. As input we have used $M_Z=(91.1888\pm
0.0044)$ GeV \cite{GLASGOW}, $m_t=(174\pm 17)$ GeV
\cite{CDF}, $\A_s=0.118\pm 0.007$ \cite{EPS} and
$\DE\A^{(5)}=0.0288\pm 0.0009$ \cite{BJKK}.  Here
$\DE\A^{(5)}$ is the contribution of the light quarks to the running of
$\A$ from low energies up to $M_Z$. It appears in the loop correction factor
$\DE r$ in Eq. (\ref{sirlin}). In addition to the LR-model parameters,
the strong coupling constant $\A_s$ and the top mass $m_t$ were allowed
to vary. The experimental values for them cited above were used as
constraints. The mass of the higgs was assumed to be between 60 and 1000
GeV  with a central value 250 GeV.

The 95\% CL results for the case with the unspecified scalar sector  are
presented in Table 1. The allowed ranges of the parameters  are slightly
larger for larger values of $\lm$.  The same holds also for the case of
minimal LR-model, results for which are presented in Table 2, and for
the minimal LR-model with a negligibly small  $W$-$W'$ mixing angle
$\zeta$, the results for which are presented in Table 3.

The experimental value of the ratio $R_b$ prefers lower values of $m_t$.
Setting $m_t=$174 GeV causes the theoretical value of $R_b$ to be two
standard deviations away from the experimental value. For example, in
the case of LR-model with unspecified scalar sector and $\lm=1$,
excluding the $R_b$-contribution lowers the minimum of $\chi^2$ to
$\chi^2_{min}=8.5$, changes the 95\% CL range of the parameter
$\DE\rho_0$ to $\DE\rho_0=(1.0\pm 3.7)\cdot 10^{-3}$,   changes the 95\%
CL range of the top mass from $m_t=152\pm 33$ GeV to $m_t=165\pm 35$ GeV
but leaves the bounds for $\xi_0$ and $M_{Z'}$ unchanged. This behaviour
can be explained by noting that the dominant $m_t$-dependence of $R_b$
comes through the term $\D_{vb}$ in Eq. (\ref{bdif}), whereas the other
observables receive a significant contribution also from
$\DE\rho\equiv\DE\rho_0+\DE\rho_t$. Here $\DE\rho_t$ is
the top mass dependent part of the Standard Model contribution
to the parameter $\rho$ and it reads, in the limit of large top mass,
$\DE\rho_t=3G_Fm_t^2/(8\sqrt{2}\pi^2)$. When $m_t$ decreases to fit
better to $R_b$, the Standard Model contribution to the parameter
$\DE\rho$ decreases allowing a larger value of $\DE\rho_0$.

The best value of the top mass was found to be, almost independently of
the model considered, to be around $m_t=150\pm 35$ GeV ( 95\% CL). We
performed also a Standard Model fit to the parameters $m_t$ and $m_H$.
We found the 68\% CL result
\beqa
m_t &=& 157^{+11}_{-11}\mbox{\ GeV}, \nonumber \\
m_H &=& 77^{+144}_{-48}\mbox{\ GeV}
\eeqa
in agreement with a recent study \cite{EFL}.

We now compare our results with those obtained in the studies \cite{LL}
and \cite{ACDDGF}. Langacker and Luo \cite{LL} used low-energy
measurements, LEP-measurements and $M_W$-measurement to fit the
parameters of the extended models. They found the 95\% CL bounds
$\xi_0=(1.8^{+6.1}_{-6.6})\cdot 10^{-3}$ and $M_{Z'}>387$ GeV in the
case of the general LR-model with $g_L=g_R$; and $M_{Z'}>857$ GeV in the
case of the  minimal LR-model with $g_L=g_R$.   Due to the increased
precision of the LEP-measurements the bounds obtained in the present
study are considerably  tighter except   for $M_{Z'}$ in the general
LR-model, for which our bound is 30 GeV lower. This is presumably due to
the larger low-energy data set used in \cite{LL}, in addition to the
experimental quantities used in the present study Langacker and Luo use
also $e^-e^+$ -annihilation data below the $Z$-pole.

Altarelli et al. \cite{ACDDGF} used the LEP-measurements and
$M_W$-measurement to fit the parameters of the extended models. In the
case of LR-model with $g_L=g_R$ and unspecified scalar sector
they found the $1\sigma$ ranges $\xi_0=(0.15\pm 1.58)\cdot 10^{-3}$ with
top mass fixed to $m_t=150$ GeV; and $\xi_0=(-0.1\pm 2.5)\cdot 10^{-3}$,
$m_t<147$ GeV when letting $\A_s$ and $m_t$ vary. The low value of the
top mass in the latter case is due the use of
$\DE\rho=\DE\rho_0+\DE\rho_t$ as a fitting parameter, causing the other
observables than $R_b$ being almost independent of the top mass.
Our results are in agreement with those cited above. The $1\sigma$ range
for $\xi_0$ in the case of LR-model with unspecified scalar sector and
$\lm=1$ is $\xi_0=(0.5\pm 1.4)\cdot 10^{-3}$. The constraint
$m_t=174\pm 17$ GeV used in our analysis raises the central value
and reduces slightly the allowed range of $\xi_0$.

In the case of minimal LR-model, one obtains from Table 2 a lower bound
also for $M_{W'}$ by using the relation
\be
M_{W'}=\frac{y}{\sqrt{2}c_w}M_{Z'},
\ee
which follows from the Eqs. (\ref{mw}) and (\ref{mz}). For example,
in the case of $\lm=1$,
$M_{W'}>740$ GeV. This bound is more restrictive than the
bound obtained from charged current data \cite{LS} in the case when
the right-handed quark mixing matrix $U^R$ is unrelated
to the left-handed CKM-matrix $U^L$,
$M_{W'}>670$ GeV (90\% CL); but is less restrictive in the
case of manifest or pseudomanifest left-right symmetry,
which implies $|U_{ij}^R|=|U_{ij}^L|$, $M_{W'}>1.4$ TeV (90\% CL).

To conclude, using the latest LEP results and the top mass constraint
$m_t=174  \pm 17$ GeV and assuming the left-right symmetric model, one
is able to constrain the $Z$-$Z'$ mixing to be smaller than 0.5 \% and
the tree level contribution $\DE\rho_0$ to the $\rho$-parameter to
be smaller than 0.6 \%. If one further assumes the LR-model with minimal
scalar sector, it is found that the mass of the heavier neutral gauge
boson should be larger than 1 TeV.

\bigskip

\noindent{\bf Acknowledgement}
Author thanks Iiro Vilja and Jukka Maalampi for a critical reading of the
manuscript.
%\newpage

%\newpage
\noindent{\bf TABLE CAPTIONS}

\noindent{\bf Table 1.} The 95\% CL neutral current
parameter limits ($\chi^2<\chi^2_{min}+4.8$ )
for the LR-model with unspecified scalar sector.

\noindent{\bf Table 2.} The 95\% CL parameter
limits for the minimal LR-model.

\noindent{\bf Table 3.} The 95\% CL lower limits of the mass of the
heavy neutral gauge boson
in the minimal LR-model with a negligible charged current mixing angle.

\newpage
\center{\bf\large Table 1.}
%\vskip 1cm
\center{\large
%\begin{table}
\begin{tabular}{ccccc}\hline
$\lambda$ & $\chi^2_{min}$ & $\Delta\rho_0$ & $\xi_0$
&$M_{Z',min}[\mbox{GeV}]$\\ \hline
1.0 & 14.5& $(2.1\pm 3.6)\cdot 10^{-3} $ & $(0.5\pm 3.1)\cdot 10^{-3} $ &
359 \\
1.1 & 14.4& $(2.1\pm 3.6)\cdot 10^{-3} $ & $(0.6\pm 3.6)\cdot 10^{-3} $ &
344 \\
1.2 & 14.3& $(2.2\pm 3.7)\cdot 10^{-3} $ & $(0.8\pm 4.1)\cdot 10^{-3} $ &
333 \\ \hline
\end{tabular}
%\caption{}
%\end{table}
}
\vskip 2.5cm
%\newpage
\center{\bf\large Table 2.}
%\vskip 1cm
\center{\large
%\begin{table}
\begin{tabular}{cccc} \hline
$\lambda$ & $\chi^2_{min}$ & $|\zeta|_{max}$ & $M_{Z',min}[\mbox{TeV}]$\\
\hline
1.0 & 15.7& $5.5\cdot 10^{-3} $ & 1.24 \\
1.1 & 15.6& $6.7\cdot 10^{-3} $ & 1.06 \\
1.2 & 15.6& $8.1\cdot 10^{-3} $ & 0.92 \\
\hline
\end{tabular}
%\caption{}
%\end{table}
}
\vskip 2.5cm
%\newpage
\center{\bf\large Table 3.}
%\vskip 1cm
\center{\large
%\begin{table}
\begin{tabular}{ccc}  \hline
$\lambda$ & $\chi^2_{min}$ & $M_{Z',min}[\mbox{TeV}]$\\ \hline
1.0 & 15.7& 1.24 \\
1.1 & 15.6& 1.06 \\
1.2 & 15.6& 0.92 \\ \hline
\end{tabular}
%\caption{}
%\end{table}
}

\end{document}